\documentclass[12pt]{article}

\usepackage{amsmath,amsthm,amssymb,amscd}
\usepackage{graphicx, float}
\usepackage[hidelinks]{hyperref}
\usepackage{natbib}
\usepackage{ds}

\usepackage[linesnumbered,ruled,vlined]{algorithm2e} % adding the algorithm box
\usepackage{tabularx} % automatic line breaks in tables
\newcolumntype{L}[1]{>{\raggedright\arraybackslash}p{#1}} % left fixed width

\usepackage{enumitem} % using indent items in enumerate
\usepackage{caption, subcaption} % using the caption for tables and figures
\usepackage{capt-of} % cross-referencing the figures in the appendix
\usepackage[flushleft]{threeparttable} % adding the footnote in a table
\usepackage{multirow}
\usepackage{booktabs}
\usepackage{eurosym}
\usepackage{xcolor,soul} % adding color fonts and highlight text

\usepackage{authblk}

\title{\bf Optimal resource allocation: Convex quantile regression approach}

\author[a,\footnote{Corresponding author. \newline
\hspace*{5mm} \textit{E-mail addresses:} \texttt{sheng.dai@utu.fi (S. Dai)}, \texttt{natalia.kuosmanen@etla.fi (N. Kuosmanen)},\\
\hspace*{34mm} \texttt{timo.kuosmanen@utu.fi (T. Kuosmanen)}, \texttt{juuso.liesio@aalto.fi (J. Liesi\"{o})}.}]{Sheng Dai}
\author[b]{Natalia Kuosmanen}
\author[a]{Timo Kuosmanen}
\author[c]{Juuso Liesi\"{o}}

\affil[a~]{Turku School of Economics, University of Turku, 20500 Turku, Finland}
\affil[b~]{ETLA Economic Research, 00100 Helsinki, Finland}
\affil[c~]{Aalto University School of Business, 02150 Espoo, Finland}
\date{\today}

\begin{document}
% revising the caption Figure XX: to Fig. XX.; Table XX: to Table XX.
\captionsetup[figure]{labelfont={bf},labelformat={default},labelsep=period,name={Fig.}}
\captionsetup[table]{labelfont={bf},labelformat={default},labelsep=period,name={Table}}

\maketitle
 
\vfill
%\begin{center}
%Preprint submitted to \emph{MS/EJOR}
%\end{center}
\vfill

\begin{abstract}
\noindent Optimal allocation of resources across sub-units in the context of centralized decision-making systems such as bank branches or supermarket chains is a classical application of operations research and management science. In this paper, we develop quantile allocation models to examine how much the output and productivity could potentially increase if the resources were efficiently allocated between units. We increase robustness to random noise and heteroscedasticity by utilizing the local estimation of multiple production functions using convex quantile regression. The quantile allocation models then rely on the estimated shadow prices instead of detailed data of units and allow the entry and exit of units. Our empirical results on Finland's business sector reveal a large potential for productivity gains through better allocation, keeping the current technology and resources fixed.
\\[5mm]
\end{abstract}
\vfill

\thispagestyle{empty}

\newpage
\setcounter{page}{1}
\setcounter{footnote}{0}
\pagenumbering{arabic}
\baselineskip 20pt

%-----------------%
%
%-----------------%

\section{Introduction}\label{sec:intro}

Efficient allocation of resources is a classical topic both in economics and operations research and management science (ORMS). In economics, the fundamental theorems of welfare economics state that competitive markets ensure an efficient allocation of resources. However, the actual allocation of resources often deviates from the ideal allocation due to various factors such as market imperfections and economic policies, resulting in the misallocation of resources (e.g., \citeauthor{Hsieh2009}, \citeyear{Hsieh2009}; \citeauthor{Restuccia2017}, \citeyear{Restuccia2017}). For instance, highly efficient enterprises may face financial constraints and lack access to credit support, while less efficient or `zombie' enterprises may occupy a significant amount of financial resources due to soft budget constraints (\citeauthor{Caballero2008}, \citeyear{Caballero2008}). Such misallocation of resources undoubtedly affects the overall economic efficiency and can at least partly explain the secular stagnation of productivity growth in Western countries since the financial crisis (see, e.g., \citeauthor{Dias2016}, \citeyear{Dias2016}; \citeauthor{Corrado2019}, \citeyear{Corrado2019}).

While the economic literature on resource reallocation emphasizes the market mechanism, in the ORMS literature, the focus is typically on centralized resource allocation problems where the central management assigns additional resources or reallocates the current resources to sub-units to achieve the maximal total output. Examples of such systems include: bank branches (\citeauthor{Ray2016}, \citeyear{Ray2016}), fire departments (\citeauthor{Athanassopoulos1998}, \citeyear{Athanassopoulos1998}), harbours (\citeauthor{Fang2016}, \citeyear{Fang2016}; \citeauthor{Lozano2011}, \citeyear{Lozano2011}; \citeauthor{Wu2016}, \citeyear{Wu2016}), supermarkets (\citeauthor{Korhonen2004}, \citeyear{Korhonen2004}; \citeauthor{Liesio2020}, \citeyear{Liesio2020}). Recently, \citet{Fujii2015} and \citet{Emrouznejad2019} extend the objective of the allocation problem to minimize the total undesirable output (e.g., CO$_2$ emissions).\footnote{
    In addition to the allocation of production resources, cost allocation across a set of production entities is another type of resource allocation problem (see, e.g., \citeauthor{Cook2005}, \citeyear{Cook2005}; \citeauthor{Dehnokhalaji2017}, \citeyear{Dehnokhalaji2017}).
}

In the previous ORMS studies, the production frontier is estimated using the deterministic data envelopment analysis (DEA) method. In the DEA resource allocation models, the DEA production frontier is given as a constraint, and the production possibility set is assumed to stay constant after the reallocation of resources. But the technical efficiency of the unit can change or be maintained due to reallocation (see, e.g., \citeauthor{Korhonen2004}, \citeyear{Korhonen2004}; \citeauthor{Lozano2011}, \citeyear{Lozano2011}). The DEA production frontier is known to be sensitive to extreme observations and outliers (\citeauthor{Dai2023}, \citeyear{Dai2023}). Furthermore, extrapolating the 100\% efficient DEA production frontier to units that operate at a low level of efficiency (e.g., less than 10\% efficiency) requires strong homoscedasticity assumptions that likely fail in the real world. Finally, the DEA allocation model does not allow the entry and exit of units, which contradicts typical managerial practices.

In practice, observed marginal products and even productivity differences between firms, which are both large and persistent even in narrowly defined industries (see, e.g., \citeauthor{Syverson2011}, \citeyear{Syverson2011}; \citeauthor{Kuosmanen2020b}, \citeyear{Kuosmanen2020b}), can be largely due to heterogeneity of production resources (e.g., labor and capital inputs). It is thus critical to take heterogeneity explicitly into account in the estimation of production function: a single production function may not capture well the marginal products of an industry consisting of a heterogeneous group of firms that differ in terms of their technological and managerial efficiency.

To address these estimation challenges, in this paper, we resort to the local estimation of production functions using convex quantile regression (CQR), which can provide a full picture of the conditional distributions for the production set (\citeauthor{Dai2020}, \citeyear{Dai2020}; \citeauthor{Kuosmanen2020}, \citeyear{Kuosmanen2020}). That is, this method enables us to estimate multiple production functions for different levels of efficiency. Following \citet{Kuosmanen2020b}, we employ 10 equidistant quantiles, representing production functions for the ten deciles of the performance distribution (i.e., ten groups representing the 0\%-10\%, 10\%-20\%, $\ldots$, 90\%-100\% levels of efficiency). Furthermore, in contrast to the full frontier, local quantiles can better predict the changes in the production set, especially for long panel data.

Another notable difference to the previous resource allocation studies is that we examine potential gains of reallocation in decentralized industries consisting of independent firms that operate in a more or less competitive environment. In this context, optimal resource allocation has not attracted much attention because it is widely believed that market competition will automatically lead to efficient allocation. However, the markets in the real world tend to be incomplete in different ways, and there is plenty of empirical evidence of misallocation in the recent economic literature (e.g., \citeauthor{Restuccia2008}, \citeyear{Restuccia2008}; \citeyear{Restuccia2017}; \citeauthor{Hsieh2009}, \citeyear{Hsieh2009}). Therefore, it is worth asking how far the current allocation of resources is from the optimal one. Using the firm-level data of Finland's business sector, we make an empirical contribution that sheds new light on this intriguing question.

Given the fact that better allocation of resources is likely to increase allocative efficiency and productivity, we develop the quantile optimal resource allocation models to quantify the potential productivity gains. Four different scenarios are designed to model efficient resource allocation. Unlike the conventional DEA allocation models, the developed quantile allocation models rely on the estimated shadow prices instead of detailed data of units and allow the entry and exit of units. The production resources can be assigned within deciles or across both between and within deciles,  making these models more closely aligned with real-world business scenarios. 

In this paper, we aim to blend the economics and ORMS perspectives on resource reallocation. The out-of-sample performance comparison and empirical application do not directly relate to centrally planned systems, but to settings in which countries and firms operate in a market economy. Modeling resource allocation as if those were centrally planned systems can provide useful insights in the context of free markets, which are not necessarily perfectly competitive. \citet{Monge2019} show that there is a significant and persistent degree of capital misallocation at the global country level. We also find that the real-world market allocation in Finnish industries is far from optimal in the application. Therefore, bridging the gap between these two perspectives can help understand resource allocation more accurately.

The rest of this paper is organized as follows. Section \ref{sec:prod} introduces the quantile production functions estimation using the CQR approach. Section \ref{sec:quantile} presents the developed quantile resource allocation models and compares their out-of-sample performance with a widely used public dataset. An empirical application to Finland's business sector is demonstrated in Section \ref{sec:bus}. Section \ref{sec:concl} concludes this paper with suggested avenues for future research.

%-----------------%
% 
%-----------------%

\section{Production functions estimation}\label{sec:prod}

The first step to model resource allocation is to empirically estimate the production functions. Consider thus the following general nonparametric production function model
\begin{equation}
    y_i = f(\bx_i) \cdot \exp(\varepsilon_i)
    \label{eq:prod}
\end{equation}
where $i = 1, \ldots, n$ denotes $n$ decision-making units (DMUs), $\bx_i \in \real^d$ is the $d$-dimensional input factors, $y_i \in \real$ is the output, and $f: \real^d \rightarrow \real$ belongs to a class of functions satisfying certain shape constraints (e.g., monotonicity, convexity, and concavity). The composite error term $\varepsilon_i$ captures the latent productivity differences across DMUs. 

The productivity differences can arise due to differences in the technology, quality of outputs $y_i$, quality of inputs (e.g., education and experience of workers, or the vintage of capital), managerial efficiency, or heterogeneous operating environment. The underlying sources of productivity differences are not of primary interest to this paper, and the main point is that we need to account for the productivity differences when estimating the marginal products. Note that the marginal products depend on $\varepsilon_i$ because $\partial y_i / \partial \bx_i = f^{\prime}(\bx_i) \cdot \exp(\varepsilon_i)$. That is, the higher the productivity level represented by $\varepsilon_i$, the higher the marginal products of inputs (e.g., labor and capital).

To take productivity differences and heterogeneity of DMUs explicitly into account, we resort to the conditional quantile production function $Q_{y_i}$ defined as (\citeauthor{Dai2020}, \citeyear{Dai2020})
\begin{equation}
    Q_{y_i} (\tau \mid \bx_i) = f(\bx_i) \cdot F^{-1}_{\varepsilon_i}(\tau)
    \label{eq:prodq}
\end{equation}
where $\tau \text{ } (0 < \tau < 1)$ is the order of quantile, and $F^{-1}$ denotes the inverse of the cumulative distribution function of $\varepsilon_i$. By controlling the parameter $\tau$, we can evaluate the potential output level obtained by $\tau \cdot 100\%$ of DMUs with the given resources $\bx$. We can also evaluate the marginal products at the relative performance level $\tau \cdot 100\%$ using $\partial Q_{y_i} (\tau | \bx_i) / \partial \bx_i$. 

We use quantiles to characterize the technology of each decile. The quantiles are indexed by $\tau = 0.05, 0.15, \ldots, 0.95$, that is, the quantiles are fitted in the middle of each decile of the performance distribution. More specifically, we identify $\tau$ that yields the best fit to DMU $i$ and use that specific $\tau$ for estimating the marginal products locally. One could use any arbitrary number of quantiles and performance groups, but ten deciles are commonly used in the previous studies (see, e.g., \citeauthor{Kuosmanen2020}, \citeyear{Kuosmanen2020}; \citeauthor{Dai2020}, \citeyear{Dai2020}; \citeauthor{Kuosmanen2020b}, \citeyear{Kuosmanen2020b}). Compared to the usual approach of applying a single production function to all DMUs, the use of ten deciles enables us to better capture the heterogeneity of DMUs.

The conditional quantile production function $Q_{y_i}$ can be characterized by a piece-wise linear function, where the intercept and slope coefficients (i.e., $\alpha$ and $\bbeta$) are constrained to satisfy the monotonicity and concavity conditions. Given the estimated coefficients $\hat{\alpha}_i^{\tau}$ and $\hat{\bbeta}_i^{\tau}$ with a specific quantile $\tau$, the fitted $\tau^\textsuperscript{th}$ quantile production function at any $\mathbf{x}$ is explicitly expressed as follow (\citeauthor{Kuosmanen2008}, \citeyear{Kuosmanen2008})
\begin{equation}
    \label{eq:fitted}
    \hat{f}^{\tau}(\mathbf{x}) =\min_{i = 1, \ldots, n} \{\hat{\alpha}_i^{\tau} + \hat{\bbeta}_i^{\tau} \mathbf{x}\} 
\end{equation}

To obtain $\hat{\alpha}_i^{\tau}$ and $\hat{\bbeta}_i^{\tau}$ in \eqref{eq:fitted} and estimated quantiles empirically, we resort to a fully nonparametric approach that does not require any assumptions about the functional form of the production function or its smoothness, but imposes the monotonicity and concavity properties implied by the weak axiom of profit maximization (\citeauthor{Varian1984}, \citeyear{Varian1984}). In practice, we need to solve the following linear programming (LP) problem for each quantile $\tau$
\begin{alignat}{2}
	\underset{\alpha,\bbeta, \varepsilon^{+},\varepsilon^{-}}{\mathop{\min }}&\, \tau \sum\limits_{i=1}^{n}{\varepsilon _{i}^{+}}+(1-\tau )\sum\limits_{i=1}^{n}{\varepsilon _{i}^{-}} &{}& \label{eq:cqr}\\ 
	\textit{s.t.}\quad
	& y_i=\alpha_i+ \bbeta_i^{'}\bx_i+\varepsilon_i^+ -\varepsilon_i^- &\quad& \forall i \notag \\
	& \alpha_i+\bbeta_i^{'}\bx_i \le \alpha_h + \bbeta_h^{'}\bx_i &{}& \forall i,h \notag \\
	& \bbeta_i\ge \bzero &{}& \forall i \notag \\
	& \varepsilon _i^{+}\ge 0,\ \varepsilon_i^{-} \ge 0 &{}& \forall i \notag
\end{alignat}

In the objective function, parameter $\tau$ assigns asymmetric weight to the negative deviations $\varepsilon_i^{-}$ and the positive deviations $\varepsilon_i^{+}$ from the quantile. The special case $\tau = 0.5$ that assigns equal weight to positive and negative deviations is referred to as the median regression. The first constraint is a linearized regression equation. The second inequality constraint imposes concavity of the production function. The third constraint guarantees the monotonicity of the production function. Our main interest is in the coefficients $\bbeta_i$, which directly indicate the marginal products of the input factors of DMU $i$, evaluated using a given quantile $\tau$. Further, problem \eqref{eq:cqr} presents the variable returns to scale (VRS) assumption for quantile production functions via the intercept term $\alpha_i$, which is a free variable. If $\alpha_i=0$, then problem \eqref{eq:cqr} becomes a constant returns to scale (CRS) model. 

As $\tau \rightarrow 1$, problem \eqref{eq:cqr} collapses to a conventional output-oriented DEA problem (\citeauthor{Kuosmanen2010a}, \citeyear{Kuosmanen2010a}). That is, the deterministic DEA production frontier is obtained as the limiting special case of problem \eqref{eq:cqr}. Specifically, we have the following LP problem as $\tau$ approaches one,
\vspace{-0.475cm}
\begin{alignat}{2}
	\underset{\alpha,\bbeta,\varepsilon^{-}}{\mathop{\min }}&\, \sum\limits_{i=1}^{n}{\varepsilon _{i}^{-}} &{}& \label{eq:cqr2}\\ 
	\textit{s.t.}\quad
	& y_i=\alpha_i+ \bbeta_i^{'}\bx_i - \varepsilon_i^- &\quad& \forall i \notag \\
	& \alpha_i+\bbeta_i^{'}\bx_i \le \alpha_h + \bbeta_h^{'}\bx_i &{}& \forall i,h \notag \\
	& \bbeta_i\ge \bzero &{}& \forall i \notag \\
	& \varepsilon_i^{-} \ge 0 &{}& \forall i \notag
\end{alignat}
where the additive DEA formulation \eqref{eq:cqr2} is slightly different from the standard DEA formulation proposed by \citet{Banker1984}, where a multiplicative form is used to calculate efficiency. But in the single-output setting, those two formulations are equivalent in the sense of efficiency estimate transformation (see Lemma 3.1 in \citeauthor{Kuosmanen2010a}, \citeyear{Kuosmanen2010a}). Note that the DEA efficiency for DMU $i$ is then obtained as the optimal solution (i.e., $\hat{\varepsilon}_i^{-}$) to problem \eqref{eq:cqr2}. By contrast to the standard DEA problem where the efficiency $\hat{\varepsilon}_i^{-}$ is separately computed for each DMU, problem \eqref{eq:cqr2} simultaneously calculates efficiency for all DMUs by minimizing the sum of $\varepsilon_i^{-}$.

While DEA has been widely used in centralized resource allocation models (see, e.g., \citeauthor{Korhonen2004}, \citeyear{Korhonen2004}; \citeauthor{Fang2016}, \citeyear{Fang2016}; \citeauthor{Liesio2020}, \citeyear{Liesio2020}), several deficiencies in the DEA allocation model are also highlighted. First, since the technical efficiency score $\sigma_i$ is estimated by the deterministic DEA production frontier, it is very sensitive to extreme observations (\citeauthor{Dai2023}, \citeyear{Dai2023}). Second, for the DMUs operating at a low level of efficiency (e.g., less than 10\% efficiency), we often require strong homoscedasticity assumptions, which likely fail in the real world. Third, to allocate production resources, the detailed data of each DMU need to be known prior. Finally, the direct modeling of entry and exit for DMUs in DEA allocation models is insufficient. 

In practice, the constrained optimization problem \eqref{eq:cqr} can be solved by any linear programming solver. In this paper, we employ the Mosek solver with the Python/pyStoNED package developed by \citet{Dai2021b}, which is freely available on GitHub,\footnote{
    The pyStoNED package, \url{https://github.com/ds2010/pyStoNED}. 
} 
and has been installed on the Statistics Finland server. We make use of the STATA-Python integration that enables us to process the firm data and run the pyStoNED code conveniently in STATA.

%-----------------%
% 
%-----------------%

\section{Resource allocation problems}\label{sec:quantile}

\subsection{Quantile allocation model}\label{sec:qmodel}

To address the notable deficiencies in DEA allocation models, we develop the following quantile allocation models to increase the robustness and provide alternatives to optimal resource allocation. In these models, our objective is to allocate the total resources at an aggregate level (e.g., an industry) denoted by vector $\mathbf{X}$ to counterfactual production plans ($\bx_i^\tau$, $y_i^\tau$) to maximize the total output of the industry under the following assumptions:

1)  Each decile of DMUs operates using the corresponding quantile production function. The quantiles $\tau = 0.05, 0.15, \ldots, 0.95$ have been estimated in Section \ref{sec:prod}. 

2) The production resources can be reallocated between DMUs; however, the total production resources of the industry remain constant.\footnote{
    It is also widely seen in practice that the social planner increases or decreases the total production resources during the reallocation. This assumption thus can be further relaxed, and we will revisit it in the out-of-sample performance comparison.
}

3) Reallocation does not influence DMUs' productive efficiency. That is, DMUs can move along the quantile production functions but not increase or decrease their efficiency.

To take the heterogeneity of DMUs and differences in their productive performance explicitly into account, we partition the sample of DMUs into 10 mutually exclusive groups based on their productive efficiency, representing the ten deciles of the performance distribution (i.e., 0\%-10\%, 10\%-20\%, $\ldots$, 90\%-100\%). Note that each group includes $n$ DMUs by construction, where $n = N/10$ and $N$ is the total sample size.

In the baseline model, the optimal resource allocation is obtained by solving the following LP problem
\begin{alignat}{2}
    \underset{\bx, y} {\mathop{\max }}\, \quad & \sum\limits_{\tau=1}^{10}{\sum\limits_{i=1}^{n}{y_i^{\tau} }} &{\quad}& \label{eq:LP}\\
    \mbox{s.t.}\quad
    & y_{i}^{\tau} \le \hat{\alpha}_h^{\tau}+ \hat{\bbeta}_h^{\tau}\bx_{i}^{\tau} &{}& \forall h,i,\tau \notag\\
    & \sum\limits_{\tau=1}^{10}{\sum\limits_{i=1}^{n}{\bx_i^{\tau} }} = \mathbf{X} &{}& \notag  \\ 
    & \bx_{i}^\tau \ge \text{0} \notag
\end{alignat}
where the objective function is set to maximize the total output of the industry. The first constraint, the technology constraint, requires that the total outputs of each quantile cannot exceed the corresponding production possibility. The second constraint, the resource constraint, is simply that total production resources used equals total supply, indicating that all production resources will be exactly allocated to the DMUs. The last constraint imposes the non-negativity of the allocated resources.

We index the counterfactual pseudo-DMUs of each quantile by $i = 1, \ldots, n$ in problem \eqref{eq:LP}. Note that these DMUs do not have any connection to the real DMUs other than that they operate using the same quantile production function. One does not need to have DMU-specific data of resources, it suffices to know the total resources of the industry $\mathbf{X}$ and the estimated coefficients $\hat{\alpha}_h^{\tau}$, $\hat{\bbeta}_h^{\tau}$ that characterize the ten quantile production functions.

In real-world business scenarios, new firms and establishments replace old ones in a process known as creative destruction (\citeauthor{Schumpeter1942}, \citeyear{Schumpeter1942}). However, the baseline formulation \eqref{eq:LP} does not allow us to model such entry or exit of firms. Furthermore, meaningful modeling of entry should somehow avoid the trivial solution where we just replicate the most productive DMU $N$ times. We thus extend the baseline model to consider the possibility of exit. Let us first introduce a binary decision variable $b_i$ that gets the value of 1 if the DMU is allocated resources, and 0 if the DMU is forced to exit. In the case where the exit is allowed, the resource allocation problem can be stated as the following mixed-integer linear programming (MILP) problem
\begin{alignat}{2}
    \underset{\bx, y, b} {\mathop{\max }}\, \quad & \sum\limits_{\tau=1}^{10}{\sum\limits_{i=1}^{n}{y_i^{\tau} }} &{\quad}& \label{eq:MIP}\\
    \mbox{s.t.}\quad
    & y_{i}^{\tau} \le \hat{\alpha}_h^{\tau}+ \hat{\bbeta}_h^{\tau}\bx_{i}^{\tau} + (1-b_i)M &{}& \forall h,i,\tau \notag  \\
    & y_{i}^{\tau} \le b_iM &{}& \forall i,\tau \notag  \\
    & \bx_{i}^\tau \le b_iM &{}& \forall i,\tau \notag  \\
    & \sum\limits_{\tau=1}^{10}{\sum\limits_{i=1}^{n}{\bx_i^{\tau} }} \le \mathbf{X} &{}& \notag \\ 
    & \bx_{i}^\tau \ge \text{0} \notag\\
    & b_i \in \{0, 1\}  \notag
\end{alignat}
where $M$ is a pre-specified large positive number. In practice, for example, the first $M$ in the second constraint can be set for each quantile with its maximum possible output, and the second $M$ in the third constraint could be $\mathbf{X}$. If $b_i=0$, then the DMU $i$ is inactive, and the technology constraints are relaxed through the second and third constraints of problem \eqref{eq:MIP}. The binary variable $b_i$ makes the problem computationally harder, but modern integer programming solvers can handle resource allocation problems with thousands of pseudo-DMUs.

In optimization problems \eqref{eq:LP} and \eqref{eq:MIP}, resources can be reallocated between DMUs that operate at different quantiles. This can be a strong assumption if performance differences mainly arise from inherent quality differences in resources. Consider, for example, the vintage of capital or the education and skills of employees. To address this issue, we can easily impose additional restrictions that allow reallocation to take place only within pseudo-DMUs of a given quantile, but not move resources from one quantile to another. In this case, problem \eqref{eq:LP} is reformulated as 
\vspace{-0.2cm}
\begin{alignat}{2}
    \underset{\bx, y} {\mathop{\max }}\, \quad & \sum\limits_{\tau=1}^{10}{\sum\limits_{i=1}^{n}{y_i^{\tau} }} &{\quad}& \label{eq:LP2}\\
    \mbox{s.t.}\quad
    & y_{i}^{\tau} \le \hat{\alpha}_h^{\tau}+ \hat{\bbeta}_h^{\tau}\bx_{i}^{\tau} &{}& \forall h,i,\tau \notag\\
    & \sum\limits_{\tau=1}^{10}{\sum\limits_{i=1}^{n}{\bx_i^{\tau} }} = \mathbf{X} &{}& \notag  \\ 
    & \sum\limits_{i=1}^{n}{\bx_i^{\tau} } = \mathbf{X}^{\tau} &{}& \forall \tau  \notag \\
    & \bx_{i}^\tau \ge \text{0} \notag
\end{alignat}

And problem \eqref{eq:MIP} can be rewritten as
\begin{alignat}{2}
    \underset{\bx, y, b} {\mathop{\max }}\, \quad & \sum\limits_{\tau=1}^{10}{\sum\limits_{i=1}^{n}{y_i^{\tau} }} &{\quad}& \label{eq:MIP2}\\
    \mbox{s.t.}\quad
    & y_{i}^{\tau} \le \hat{\alpha}_h^{\tau}+ \hat{\bbeta}_h^{\tau}\bx_{i}^{\tau} + (1-b_i)M &{}& \forall h,i,\tau \notag  \\
    & y_{i}^{\tau} \le b_iM &{}& \forall i,\tau \notag  \\
    & \bx_{i}^\tau \le b_iM &{}& \forall i,\tau \notag  \\
    & \sum\limits_{\tau=1}^{10}{\sum\limits_{i=1}^{n}{\bx_i^{\tau} }} \le \mathbf{X} &{}& \notag \\ 
    & \sum\limits_{i=1}^{n}{\bx_i^{\tau} } \le \mathbf{X}^{\tau} &{}& \forall \tau  \notag \\
    & \bx_{i}^\tau \ge \text{0} \notag\\
    & b_i \in \{0, 1\}  \notag
\end{alignat}
where $\mathbf{X}^\tau$ is a vector of total resources assigned to quantile $\tau$ in the observed allocation. Note that the resource allocation models \eqref{eq:LP}-\eqref{eq:MIP2} are solved by the Gurobi solver within the Python environment. 

To summarize, we develop the quantile allocation models using the following four alternative sets of constraints:
\begin{itemize}
    \item Problem \eqref{eq:LP}: Maximize output allowing reallocation both between and within deciles, no exit allowed.
    \item Problem \eqref{eq:LP2}: Maximize output allowing reallocation only within deciles, no exit allowed.
    \item Problem \eqref{eq:MIP}: Maximize output allowing reallocation both between and within deciles, exit allowed.
    \item Problem \eqref{eq:MIP2}: Maximize output allowing reallocation only within deciles, exit allowed.
\end{itemize}

Note that the output of the optimized allocation will be higher if reallocation between deciles is possible or if the exit of DMUs is allowed because in such cases the constraints of the resource allocation problem are less restrictive. 

\subsection{Out-of-sample performance comparison}

To gain an intuition on the difference between DEA and quantile resource allocation models, we resort to a panel of 38 OECD countries in the years 2015--2019 to compare the prediction power and optimal allocated output between these two approaches. This dataset was collected from the Penn World Table 10.01 (PWT 10.01) (\citeauthor{feenstra2015}, \citeyear{feenstra2015}).

In particular, the output ($y$) in each country and year is measured with the variable \textit{cgdpo}, the production-side real GDP at current purchasing power parities (PPPs) (in millions of 2017 US\$) from the PWT 10.01. The labor ($L$) and capital ($K$) inputs in each country/year are the number of persons engaged (in millions) and capital stock at current PPPs (in millions 2017US\$) (i.e., \textit{emp} and \textit{cn} in the PWT 10.01, respectively). 

We first validate the quantile approach by comparing its out-of-sample predictive power with the standard DEA approach. Suppose we observe the nearest quantile of country $i$ in the year 2015 and $L$, $K$ in the year 2016,\footnote{
The nearest quantile is determined by the difference between $\hat{\varepsilon}_i^+$ and $\hat{\varepsilon}_i^-$ solved in problem \eqref{eq:cqr}. See \citet{Kuosmanen2020b} for further discussion on searching the nearest quantile. 
} 
and taking the estimated quantiles as given, we then predict output $y^*$ of country $i$ in the year 2016. We compare it to the actual output $y$ in the year 2016 and calculate the out-of-sample prediction accuracy of the CQR approach as measured by mean squared error (MSE), $\E[\sum_i^n(y_i^*-y_i)^2]$. Similarly, we calculate the MSE of the standard DEA approach to resource allocation, where one estimates the DEA production frontier and the predicted $y^*$ is calculated, assuming the DEA efficiency score of the previous year would continue to hold next year. 

Table \ref{tab:mse} reports the MSE values of the output prediction using the CQR and DEA approaches. We observe that the quantiles perform better than the conventional DEA approach, with a smaller MSE value each year. This indicates that local quantiles can better predict short-term changes in the output, resulting from natural reallocation of resources at the macro level.
\begin{table}[H]
  \centering
  \caption{Comparison of the out-of-sample MSE of CQR and DEA.}
    \begin{tabular}{llrrrrr}
    \toprule
          && 2016  & 2017  & 2018  & 2019  & \multicolumn{1}{l}{Average} \\
    \midrule
    CQR   && 1569  & 2459  & 2170  & 677   & 1719 \\
    DEA   && 1681  & 3708  & 2379  & 818   & 2147 \\
    \bottomrule
    \end{tabular}%
  \label{tab:mse}%
\end{table}%

We next compare optimal total outputs estimated by the DEA and quantile allocation models. To illustrate, we consider the following two scenarios in the DEA allocation models: 1) the aggregate inputs ($L$ and $K$) in DEA1 are allowed to increase at most by 1\% from their observed values, and the change in input of each country can decrease at most 10\% and increase no more than 30\% from its observed values (cf. \citeauthor{Korhonen2004}, \citeyear{Korhonen2004}; \citeauthor{Liesio2020}, \citeyear{Liesio2020}). 2) the aggregate inputs in DEA2 during the reallocation remain constant, which is more comparable with the developed quantile allocation models. For fair comparisons, we relax the second assumption in Section \ref{sec:qmodel} and allow the aggregate inputs to increase at most by 1\%. The aggregate inputs $\mathbf{X}$ in quantile allocation models then become $1.01\mathbf{X}$ (e.g., the second constraint of problem \eqref{eq:LP} is replaced by $\sum_{\tau=1}^{10}{\sum_{i=1}^{n}{\bx_i^{\tau} }} = 1.01\mathbf{X}$).
 \begin{figure}[H]
    \centering
    \includegraphics[width=0.75\textwidth]{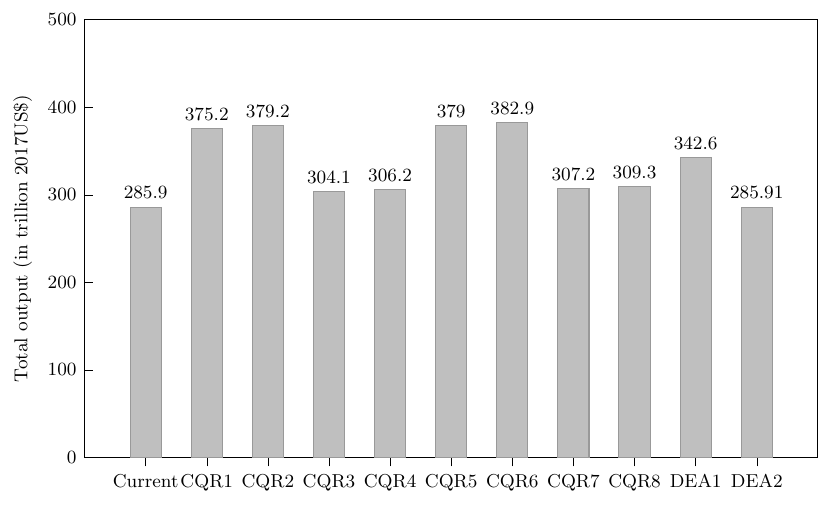}
    \hspace*{1em}\begin{minipage}[c]{0.85\textwidth}
    \footnotesize{Current: Current allocation; CQR1: problem \eqref{eq:LP}; CQR2: problem \eqref{eq:MIP}; CQR3: problem \eqref{eq:LP2}; CQR4: problem \eqref{eq:MIP2}; CQR5: problem \eqref{eq:LP} with $1.01\mathbf{X}$; CQR6: problem \eqref{eq:MIP} with $1.01\mathbf{X}$; CQR7: problem \eqref{eq:LP2} with $1.01\mathbf{X}$; CQR8: problem \eqref{eq:MIP2} with $1.01\mathbf{X}$.}
    \end{minipage}
    \caption{Comparison of optimal total output with different allocation models.}
    \label{fig:fig1}
\end{figure}

The quantile allocation models yield higher total output than the considered DEA allocation models (see Fig.~\ref{fig:fig1}). When aggregate inputs are allowed to increase, the optimal total output becomes larger in both quantile and DEA allocation models. When the resources can be allocated between and within deciles (CQR1 and CQR2), the optimal total output gets bigger than that within only deciles (CQR3 and CQR4). Furthermore, when increasing the percentage of the aggregate inputs for both DEA1 and CQR5-CQR8 in additional experiments, we observe that the more allowed increase in aggregate inputs leads to a greater increase in optimal output. However, compared with quantile models, DEA2 only has a tiny increase in optimal output if we keep the total used inputs unchanged. Overall, in all considered scenarios, more efficient allocation of resources across the OECD countries could increase economic well-being.

%-----------------%
% 
%-----------------%

\section{Application to Finland's business sector}\label{sec:bus}

\subsection{Data and variables}

In this section, we apply the proposed quantile allocation models to Finland's business sector.\footnote{
    This application is based on the project commissioned by the Prime Minister's Office of Finland. See the technical report \citet{Kuosmanen2022b} for more detailed discussion and evidence on other industries.
}
Considering that the estimation of marginal products of input factors is computationally demanding, we focus on examining the following three selected industries in the years 2005, 2012, and 2018, which yields a total of 9 distinct samples
	\begin{itemize}
		\item Manufacture of basic metals (C26; the Finnish TOL 2008 industry classification)
		\item Construction of residential and non-residential buildings (F41200)
		\item Computer programming activities (J62010) 
	\end{itemize}
	
We use the Financial Statement Data Panel of Statistics Finland, which contains firm-level accounting data covering exhaustively all enterprises in almost all industries. The output is measured by the value added (thousand \officialeuro), the labor input is measured by the number of employees (in the full-time equivalent units), and the capital input is measured by the fixed assets (thousand \officialeuro). 

To keep the sample size manageable in industries consisting of large numbers of small firms, we exclude firms with less than one employee in industry C26, less than five employees in industry J62010, and less than ten employees in industry F41200. The observations with missing values are also excluded from the sample. The number of firms in each sample is then summarized in Table~\ref{tab:sample}. Furthermore, all nominal values are deflated to the constant prices of the year 2010 using the GDP deflator of Statistics Finland.
\begin{table}[H]
  \centering
  \caption{Number of firms in each sample.}
    \begin{tabular}{p{18em}p{3em}p{3em}p{3em}}
    \toprule
          	& 2005  & 2012  & 2018 \\
    \midrule
    Manufacture of basic metals (C24)     & 134   & 127   & 100 \\
    Construction of building (F41200)	  & 642   & 783   & 1883 \\
    Computer programming (J62010)	      & 585   & 586   & 807 \\
    \bottomrule
    \end{tabular}%
  \label{tab:sample}%
\end{table}%

While we have excluded the firms with a few employees and removed the observations with missing values, the huge heterogeneity in firms' input factors can still be observed in the production function estimation. The application to Finland's business sector does not directly involve centrally planned systems, but firms operate in a market economy, which might be a monopolistic competition market. Further, DEA is a special case of the CQR model when $\tau \rightarrow 1$. In the following discussion, we thus merely apply the developed quantile allocation models to Finland's business sector.

\subsection{Marginal products}

Table~\ref{tab:mp} reports the averages of the estimated marginal products and unit costs for the selected industries in 2005, 2012, and 2018. Consider first the manufacture of basic metals (C24). For convenience, we also report the ratios of these two averages: recall that the ratio is equal to one in the competitive equilibrium of price-taking profit-maximizing firms, whereas the ratio less than one indicates under-utilization of the resource and the ratio greater than one points towards over-use of the resource. In the manufacture of basic metals, the ratios of the marginal products and unit costs are relatively close to one in 2005. The allocation of labor input remains relatively good in 2012 and 2018, but the capital input appears more problematic in this industry: the estimated marginal products point towards over-capacity in 2012, which has changed to under-capacity in 2018.  
\begin{table}[H]
  \centering
  \caption{Estimated marginal products, unit costs, and their ratios.}
    \begin{tabular}{rlrrrrrrr}
    \toprule
    \multicolumn{1}{l}{\multirow{2}[4]{*}{}} & \multicolumn{1}{l}{\multirow{2}[4]{*}{}} & \multicolumn{3}{c}{Labor (\officialeuro / worker)} &       & \multicolumn{3}{c}{Capital} \\
\cmidrule{3-5}\cmidrule{7-9}          &       & 2005  & 2012  & 2018  &       & 2005  & 2012  & 2018 \\
    \midrule
    \multicolumn{1}{l}{C24} & Unit costs & 41361 & 42368 & 43336 &       & 0.56  & 0.46  & 0.39 \\
          & Marginal product & 46033 & 45576 & 40069 &       & 0.6   & 0.28  & 0.58 \\
          & Unit costs/marginal product & 0.9   & 0.93  & 1.08  &       & 0.94  & 1.68  & 0.66 \\
          &       &       &       &       &       &       &       &  \\
    \multicolumn{1}{l}{F41200} & Unit costs & 37698 & 40503 & 39658 &       & 0.89  & 0.67  & 0.79 \\
          & Marginal product & 60051 & 60048 & 59510 &       & 0.49  & 0.35  & 0.46 \\
          & Unit costs/marginal product & 0.63  & 0.67  & 0.67  &       & 1.79  & 1.9   & 1.72 \\
          &       &       &       &       &       &       &       &  \\
    \multicolumn{1}{l}{J62010} & Unit costs & 53593 & 56025 & 53328 &       & 0.91  & 0.99  & 0.91 \\
          & Marginal product & 73699 & 75902 & 74706 &       & 0.33  & 0.37  & 0.07 \\
          & Unit costs/marginal product & 0.73  & 0.74  & 0.71  &       & 2.75  & 2.68  & 13.55 \\
    \bottomrule
    \end{tabular}%
  \label{tab:mp}%
\end{table}%

Next, consider the construction of residential and non-residential buildings (F41200). In this industry, the marginal product of labor is, on average, considerably lower than the unit cost in all three years considered. This points towards under-employment. In contrast, the marginal product of capital falls short of the average unit cost in all years, suggesting the capital intensity is higher than optimal in this industry. There are several possible explanations for this finding. Note that this industry has experienced major growth over the past decades, especially in the urban centers of Finland. As a result, many firms have a shortage of skilled workers, which can contribute to the excessive capital intensity. On the other hand, a large proportion of employees in this industry are foreign workers whose bargaining power in wage negotiations can be lower than that of native employees. Finally, the largest construction firms tend to outsource a large proportion of manual labor to subcontractors, which can bias the functional distribution of the factor shares as the outsourced labor is treated as an intermediate input.    

Regarding the computed programming industry (J62010), similar to the construction industry, the marginal product of labor is, on average, considerably lower than the unit cost in all three years considered, whereas the marginal product of capital is lower than the average unit cost in all years, suggesting the capital intensity is higher than optimal also in this industry. This industry has been the fastest-growing export industry in Finland and has also had a shortage of skilled programmers.    

\subsection{Optimal allocations}

Consider the baseline allocation problem without exit possibility (i.e., problem \eqref{eq:LP}). The optimal solution provides information on how large a proportion of resources to allocate to each decile of the performance distribution, but the allocation to firms within the group is completely immaterial because each firm of a given group is assumed to operate with the same technology. Typically, all firms at the given deciles receive exactly the same resources in the optimal solution, except for leftover resources that cannot be equally divided. 

Table~\ref{tab:op} reports the optimal shares of labor, capital, and output for each decile in the no exit scenario of the year 2005. In the manufacture of basic metals industry (C24), virtually all resources are concentrated on the most efficient deciles 1, 2, 3, and 5. All other quantiles only get the minimal resources (rounded to zero in Table~\ref{tab:op}) to keep up firms in operation. Interestingly, not all resources are allocated to the most productive firms, in fact, the largest share is allocated to quantile 3, which represents 70-80\% of the performance distribution. Note further that the fifth quantile (50-60\% of the performance distribution) operates more labor-intensively than other quantiles in the optimal solution.
\begin{table}[H]
  \centering
  \caption{The shares of labor, capital, and output for each decile in the optimal allocation for the year 2005 in the no exit case.}
  \footnotesize{
    \begin{tabular}{lrrrrrrrrrrr}
    \toprule
    \multicolumn{1}{c}{\multirow{2}[4]{*}{Decile}} & \multicolumn{3}{c}{C24} &       & \multicolumn{3}{c}{F41200} &       & \multicolumn{3}{c}{J62010} \\
    \cmidrule{2-4}\cmidrule{6-8}\cmidrule{10-12}   & Labor & Capital & Output&       & Labor & Capital & Output   &       & Labor & Capital & Output \\
    \midrule
    1 (90-100\%)    & 23    & 27    & 26    &       & 41    & 7     & 52    &       & 36    & 43    & 51 \\
    2 (80-90\%)    & 18    & 29    & 25    &       & 19    & 73    & 24    &       & 51    & 42    & 44 \\
    3 (70-80\%)    & 35    & 36    & 36    &       & 4     & 1     & 2     &       & 3     & 15    & 3 \\
    4 (60-70\%)    & 0     & 0     & 0     &       & 24    & 15    & 19    &       & 1     & 0     & 1 \\
    5 (50-60\%)    & 24    & 8     & 13    &       & 2     & 1     & 1     &       & 1     & 0     & 0 \\
    6 (40-50\%)    & 0     & 0     & 0     &       & 2     & 0     & 1     &       & 1     & 0     & 0 \\
    7 (30-40\%)    & 0     & 0     & 0     &       & 1     & 0     & 0     &       & 1     & 0     & 0 \\
    8 (20-30\%)    & 0     & 0     & 0     &       & 2     & 0     & 1     &       & 1     & 0     & 0 \\
    9 (10-20\%)    & 0     & 0     & 0     &       & 2     & 2     & 1     &       & 1     & 0     & 0 \\
    10 (0-10\%)    & 0     & 0     & 0     &       & 2     & 1     & 1     &       & 2     & 1     & 0 \\
    \bottomrule
    \end{tabular}%
    }
  \label{tab:op}%
\end{table}%

For the construction of residential and non-residential buildings industry (F41200), all deciles receive at least 1-2 percent of labor resources in the optimal allocation, but most resources are assigned to the most productive deciles 1-4. Interestingly, the most productive decile operates using a relatively labor-intensive technology in the optimal allocation, whereas the second decile of the productivity distribution takes a more capital-intensive approach.

Finally, Table~\ref{tab:op} also presents the analogous results for the computer programming industry (J62010). All deciles receive at least one percent of labor resources in the optimal allocation, but most resources are assigned to the most productive deciles 1-3. The most productive decile operates using a relatively capital-intensive technology in the optimal allocation, whereas the second decile is assigned more than half of the total labor resources.  

In conclusion, these three industries illustrate that it is beneficial to concentrate more resources on the top deciles of the productivity distribution. However, it is not necessarily optimal to allocate all resources to the most productive firms. The returns to scale in production seem to drive this result. Our empirical estimation of the quantile production function allows for variable returns to scale: there can be first increasing returns to scale that turn to decrease returns to scale after the most productive scale size has been reached. Note that the returns to scale properties are not imposed but are estimated in a data-driven fashion: we only assume monotonicity and concavity of the quantile production functions. 

The allocation results also illustrate that the optimal capital intensity can differ across different productivity levels. There can be room for more capital-intensive and more labor-intensive clusters of firms even within the same narrowly defined industry. 

\subsection{Productivity gains}

In this section, we assess the efficiency of the current allocation relative to the optimal allocations. Let $y^*$ be the maximum output that the industry could produce with the given input resources if optimally allocated across firms. If $y$ is the current level of output, then the allocative efficiency of the industry can be measured as $y/y^* \times 100\%$. The inverse of the allocative efficiency indicates the potential increase in output that could be achieved by improving the allocative efficiency of the industry. The maximum outputs have been computed using the four quantile allocation models \eqref{eq:LP}-\eqref{eq:MIP2} as explained in Section \ref{sec:qmodel}.  

Given the fact that the real-world allocation is far apart from the optimal one, we thus consider another benchmark to compare the real-world allocation: how bad is the real allocation compared to just purely random allocations? We design the following Algorithm \ref{alg1} to implement the random allocations, which are computed using the same quantile production functions as those used in solving the optimal allocations. Given the total of labor ($L^{\mathrm{total}}$), the total of capital ($K^{\mathrm{total}}$), and the estimated coefficients ($\hat{\alpha_i}^{\tau}$ and $\hat{\bbeta}_i^{\tau}$), we resort to Algorithm \ref{alg1} to obtain the average and median of the 1000 times simulated random allocations. 

\vspace{1em}
\begin{algorithm}[H]\label{alg1} 
  \KwData{$L^{\mathrm{total}}$, $K^{\mathrm{total}}$, $\hat{\alpha_i}^{\tau}$, $\hat{\bbeta}_i^{\tau}$, and $N$}
  $\mathrm{out} = 0$ and $n = \lceil N/10 \rceil$\;
  \While{$\mathrm{out} < 1000$}{
    \For{$\tau = 0;\ \tau < 10;\ \tau = \tau + 1$}{
    Generate two random numbers: $l \in \real^n \sim U[0, 1]$, $k \in \real^n \sim U[0, 1]$\;
    Calculate allocated resources for each pseudo-DMU $j$: $L_j = L^{\mathrm{total}} \times l_j/\sum\limits_{j=1}^{n}l_j$, $K_j = K^{\mathrm{total}} \times k_j/\sum\limits_{j=1}^{n}k_j$, and let $\mathbf{x} = \{L_j, K_j\}$\;
    Compute $\hat{y}_j^{\tau}$ by using equation \eqref{eq:fitted} with $\mathbf{x}$, $\hat{\alpha_i}^{\tau}$, and $\hat{\bbeta}_i^{\tau}$\;
    }
    Calculate the total output of an industry: $Y = \sum\limits_{j=1}^{n}\sum\limits_{\tau=1}^{10}\hat{y}_j^{\tau}$\;
    $\mathrm{out} = \mathrm{out} + 1$\;
 }
 Compute the average and median output: $\overline{Y} = \frac{1}{1000} Y$, $\Tilde{Y} = \mathrm{Median}\{Y_1, \ldots, Y_{1000}\}$\;
 \KwResult{$\overline{Y}$ and $\Tilde{Y}$}
\caption{Random resource allocation.}
\end{algorithm}
\vspace{2em}

\begin{figure}[H]
    \centering
    \includegraphics[width=\textwidth]{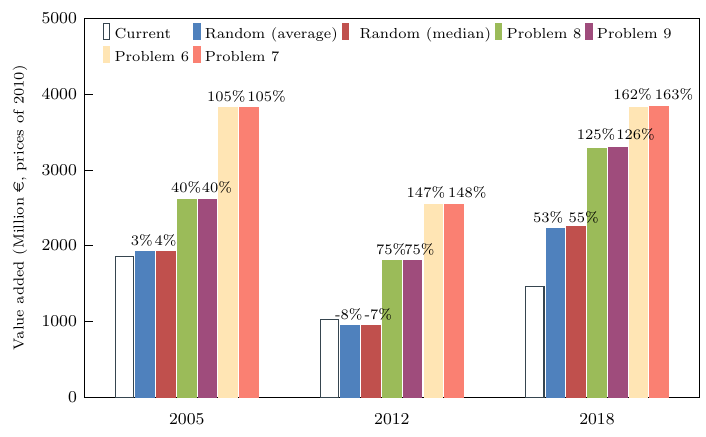}
    \caption{C24: Value added and the potential percentage increase.}
    \label{fig:C24}
\end{figure}

Fig.~\ref{fig:C24} depicts the value added of the basic metals industry (C24) in the current allocation, random allocations, and the four optimized allocations in the years 2005, 2012, and 2018. We also report the potential percentage change of output through reallocation relative to the current allocation. We find that the basic metals industry achieved similar output as the random allocations in 2005 and 2012, but fell notably short of the random allocations in 2018. Optimizing the allocation by keeping the total resources of each decile fixed would already yield a substantial increase in output, ranging from 40 percent in 2005 up to 125 percent in 2018. The potential benefit of reallocation further increases if we allow reallocating capital and labor across quantiles.

\begin{figure}[H]
    \centering
    \includegraphics[width=\textwidth]{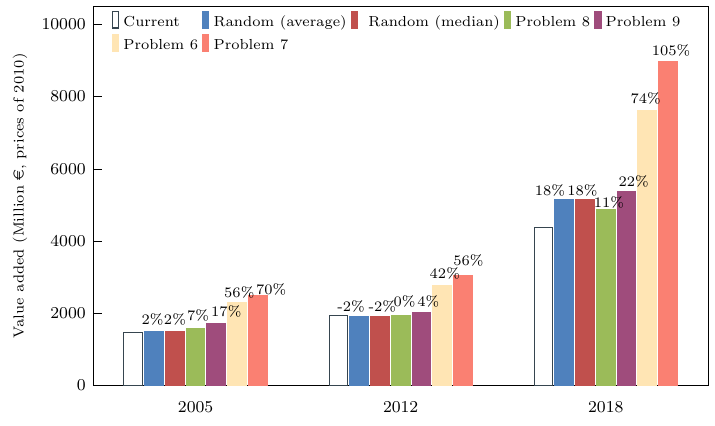}
    \caption{F41200: Value added and the potential percentage increase.}
    \label{fig:F41200}
\end{figure}

The construction industry (F41200) is also competitive with the random allocations in 2005 and 2012, but does not reach its potential in 2018 (Fig.~\ref{fig:F41200}). Optimizing the allocation by keeping the total resources of each decile fixed would yield a relatively modest increase in output, ranging from zero to 11 percent in the no exit scenarios and from four to 22 percent when the forced exit is allowed. If reallocation of capital and labor between quantiles is allowed, the potential benefits of reallocation increase considerably, ranging from 42 to 105 percent, depending on the year and whether forced exit is allowed or not.  

\begin{figure}[H]
    \centering
    \includegraphics[width=\textwidth]{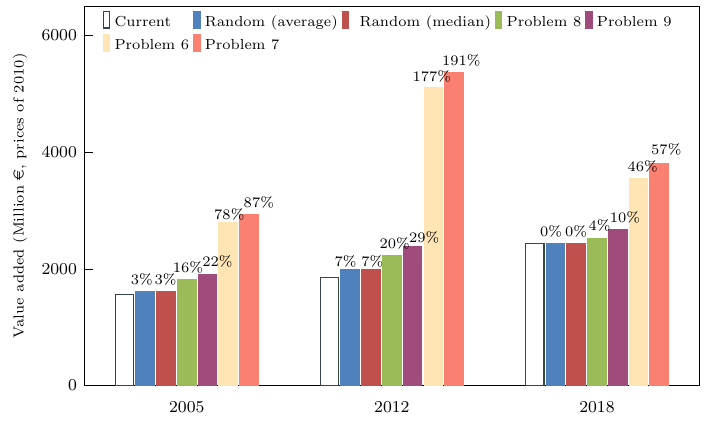}
    \caption{J62010: Value added and the potential percentage increase.}
    \label{fig:J62010}
\end{figure}

The computer programming industry (J62010) is relatively competitive with random allocations in all years (Fig.~\ref{fig:J62010}). Optimizing the allocation within deciles would yield a notable increase in output, ranging from four to 20 percent in the no exit scenarios and from 10 to 29 percent when exit is allowed. When the reallocation of capital and labor across quantiles is considered, the benefits of reallocation sharply increase, especially in the year 2012.

Having discussed the optimal allocations in detail, we next turn to explore the allocative efficiency of the current allocation relative to the optimal allocation in the case where reallocation is possible within the deciles, but reallocation between deciles is not allowed. Table \ref{tab:prod} reports the estimated allocative efficiency as a percentage of the value added in the current allocation relative to the value added in the optimal allocation for the 16 selected industries in the years 2005, 2012, and 2018, computed both with and without the exit possibility. Overall, the allocation turns out to be relatively inefficient in almost all scenarios. Furthermore, for the scenario where resources can be reallocated within and between deciles, the allocative efficiency decreases further compared to the case where reallocation is only allowed within the deciles, as expected.
\begin{table}[H]
  \centering
  \caption{Allocative efficiency (\%) of three selected industries.}
    \begin{tabular}{rlrrrrrrrr}
    \toprule
    \multicolumn{1}{c}{\multirow{2}[4]{*}{Model specification}} & \multicolumn{1}{c}{\multirow{2}[4]{*}{Industry}} & \multicolumn{2}{c}{2005} &       & \multicolumn{2}{c}{2012} &       & \multicolumn{2}{c}{2018} \\
\cmidrule{3-4}\cmidrule{6-7}\cmidrule{9-10}          &       & No exit & Exit  &       & No exit & Exit  &       & No exit & Exit \\
    \midrule
    \multicolumn{1}{p{7.5em}}{within deciles} & C24   & 71.2  & 71.3  &       & 57.2  & 57    &       & 44.4  & 44.2 \\
          & F41200 & 93.5  & 85.4  &       & 100   & 95.9  &       & 89.9  & 81.7 \\
          & J62010 & 85.9  & 82.1  &       & 83.1  & 77.7  &       & 96.6  & 90.6 \\
          &        &       &       &       &       &       &       &       &  \\
    \multicolumn{1}{p{7.5em}}{both within and between deciles} & C24   & 48.7  & 48.7  &      & 40.5  & 40.3  &       & 38.2  & 38 \\
          & F41200 & 64.1  & 58.9  &       & 70.4  & 63.9  &       & 57.6  & 48.9 \\
          & J62010 & 56.1  & 53.5  &       & 36.2  & 34.4  &       & 68.3  & 63.7 \\
    \bottomrule
    \end{tabular}%
  \label{tab:prod}%
\end{table}%

It is interesting to note that the possibility of force exit has only a marginal impact on the optimal allocation in most industries. This suggests that a large majority of the observed firms are viable in the optimal allocation, the biggest productivity gains could be achieved by better allocation of resources between existing firms. Only in the construction industry, the forced exit of the least efficient firms would seem to yield notable productivity gains in all three years considered. 

While the deciles can differ in terms of quality of resources (e.g., vintage of capital, skills, and education of employees), most likely at least some reallocation of resources between the deciles should be feasible. If we interpret the results of within deciles as an upper bound and those of both within and between deciles as a lower bound for allocative efficiency, then the level of allocative efficiency appears to be in the ballpark range of 40-70 percent in the manufacturing industry. That is, there is enormous potential for productivity growth at the industry level through better allocation of resources, which does not require more resources, technical progress, or any efficiency improvement at the firm level.    

%-----------------%
% 
%-----------------%

\section{Conclusions}\label{sec:concl}

It is widely held that free-market competition automatically ensures efficient allocation of resources. However, recent economic literature has found evidence of systematic misallocation of labor and capital in many countries (e.g., \citeauthor{Foster2001}, \citeyear{Foster2001}, \citeyear{Foster2008}; \citeauthor{Hsieh2009}, \citeyear{Hsieh2009}). In the spirit of the previous literature but developing more robust quantile allocation models, we observe large and persistent allocative inefficiencies in three relatively homogenous industries in Finland.

The comparison of the estimated marginal products and the average unit costs points towards notable capital bias in these three industries and the years considered. The average unit costs of capital exceed the marginal product, and capital resources remain inefficiently used in Finnish industries. In contrast, the marginal product of labor typically exceeds that of the average unit cost, which suggests that it would be socially optimal for most firms to hire more employees. 

The current allocation of resources is barely competitive with random allocations and is a far cry from the optimal allocation. The industries examined achieve only about a half of the potential output that could be produced with the same labor and capital resources, and by using the same technology at the constant level of productivity, if only the resources were more efficiently allocated across the observed firms. 

The examination of the optimal allocations suggests that it would be more efficient to concentrate resources to the top deciles of the performance distribution to benefit from the economies of scale. However, it is not necessarily optimal to assign all resources or even the largest share of resources to the most productive firms. There can be viable niche firms that can combine a more capital-intensive or a more labor-intensive profile with a highly productive scale size. 

It would be important to gain a better understanding of how government policy could help stimulate and steer firms to achieve better allocation of resources to improve aggregate productivity. On the one hand, misallocation can result from a lack of competition in fragmented local markets, including the labor markets for employees with highly specific skills. On the other hand, competition policy might present obstacles to more efficient coordination between different firms in the innovation ecosystems and value chains. Based on the present study, one cannot conclude if more competition or more coordination would be needed. We leave this as an interesting challenge for further research. 

%-----------------%
% 
%-----------------%

\section*{Acknowledgments}\label{sec:ack}

Sheng Dai gratefully acknowledges financial support from the OP Group Research Foundation [grant no.~20230008] and the Turku University Foundation [grant no.~081520].

%-----------------%
% 
%-----------------%

\baselineskip 12pt
\bibliographystyle{dcu}
\bibliography{References.bib}

%-----------------%
% 
%-----------------%

%\clearpage
%\newpage
%\baselineskip 20pt
%\section*{Appendix}\label{sec:app}

%-----------------%
% 
%-----------------%

%\renewcommand{\thesubsection}{\Alph{subsection}}
%\renewcommand{\thetable}{A\arabic{table}}
%\setcounter{table}{0}
%\renewcommand{\theequation} {A.\arabic{equation}}
%\setcounter{equation}{0}

\end{document}